\documentstyle[prb,aps]{revtex}
\begin{document}
\twocolumn[\hsize\textwidth\columnwidth\hsize\csname@twocolumnfalse%
\endcsname
\title{The Ground State of a Two-dimensional Lattice
System with a Long-range Interparticle Repulsion. Effective Lowering of
Dimension.}
\author{A.~A.~Slutskin, V.~V.~Slavin, and H.~A.~Kovtun.}
\address{Institute for Low Temperature Physics and  Engineering, 47 Lenin
Ave., Kharkov, Ukraine.}
\date*
\maketitle
\draft
\begin{abstract}
It has been shown that effective lowering of dimension underlies
ground-state space structure and  properties of two-dimensional lattice
systems with a long-range interparticle repulsion.  On the basis of this
fact a rigorous general procedure has been developed to describe the
ground state of the systems.
\end{abstract}
]
The two-dimensional (2D) lattice systems with a long-range interparticle
repulsion (LSLRIR) are of great interest as they have important physical
applications. One of them is the adatom systems \cite{Pokrovsky} with a
sufficiently strong interaction between the particles and the substrate.
Another field to apply the 2D LSLRIR model is a `frozen' electron phase in
2D or layered narrow-band conductors with a {\em long-ranged}
electron-electron repulsion.  This state results from a {\em suppression}
of a narrow-band electrons' tunneling between host-lattice sites which is
produced by the mutual electron repulsion \cite{FNT93_UFN95} and,
therefore, differs principally from a Wigner crystal. (It arises if $t<
\delta u = (a/\overline r)\overline u$, where $t$ is the electron
bandwidth, $\delta u$ is the typical change in the energy of a narrow-band
electron as it hops between the host-lattice sites, $a$ is the
host-lattice spacing, $\overline r$ is the mean electron separation,
$\overline u$ is the mean Coulomb energy per electron).  Layered
conductors such as cuprates and polycrystal boundaries of the non-linear
electroceramic materials \cite{Heywang} as well as art 2D conductors
\cite{Pepper} appear to be most favorable for the  electron `freezing'.

At $t\ll \delta u$ the  2D - `frozen' - electron phase
ground-state space structure is much the same as that of the adatom
systems. As far as we know, even in this limit only the one-dimensional
(1D) LSLRIR have been studied adequately (ground state
\cite{Hubbard,Bak,Synay}, thermodynamics \cite{SS}). As to the 2D LSLRIR,
neither its ground state nor thermodynamics have been explicated. Here we
develop a rigorous procedure to describe, in the zero-bandwidth limit, the
ground state of the 2D LSLRIR with an {\em isotropic} pair potential of
the interparticle repulsion, $v(r)$ ($r$ is a distance between interacting
particles). The basis of our consideration is {\em zero-temperature
effective lowering of dimension} in the 2D LSLRIR, which we have found to
take place irrespective of a form of $v(r)$ (within the limits of weak and
physically reasonable restrictions)) and for any filling factor, $\rho =
N/{\cal N}$ ($N$ and $\cal N$ are the total numbers of the particles and
host-lattice sites respectively, $N,{\cal N}\rightarrow \infty$). We take
the term to mean that the ground state of the 2D LSLRIR is an effective 1D
LSLRIR whose ``particles'' are strip-like periodic structures on the 2D
host lattice.  This enables to describe the ground state analytically,
using results of the 1D LSLRIR theory \cite{Hubbard,Bak,Synay}.

In the limiting case under consideration the Hamiltonian, $\cal H$, of the
2D LSLRIR is of the form

\begin{equation}
\label{H}
{\cal H}={\frac12}\sum_{\vec r\neq\vec {r'}}
{v\left(|\vec r-\vec {r'}|\right)n(\vec r)n(\vec {r'})}\,,
\end{equation}

\noindent where $\vec r$ are radius vectors of sites of the host lattice;
it is assumed to be a {\em triangular} one with $a = 1$; $n(\vec r)=0
\text { or } 1$ is the number of particles at a given site; the sum is
taken over the whole host-lattice; function $v(r)$ is assumed to decrease
faster than $r^{-2}$ as $r\rightarrow\infty$ and to be changed
substantially over distances $\sim \overline r$.

It is reasonable to build up the 2D LSLRIR ground state description around
the simplest 2D structures which are stable triangular 2D crystals with one
particle per cell and $\rho = \rho_q\equiv 1/q^2$ (i.e. their primitive
translation vectors have a form $q\vec a$, where $\vec a$ are those of the
host lattice). We will call them ``$q$-crystals''. The stability of the
$q$-crystals is evident, since it is a triangular lattice that realizes the
absolute minimum of the energy of mutual particle repulsion if the
particles are free to move.  With a view to formulating the general
approach to the problem, we start with consideration of small vicinities
of $\rho_q$.  As infinitesimal transformations of the system with the
Hamiltonian (\ref{H}) are impossible, a small change in $N$ results only
in formation of {\em isolated defects} in a $q$-crystal, the space
structure of the defects essentially depending on whether they arise on an
increase or on a decrease in $N$. This fact can formally be expressed by
the identity

\begin{eqnarray}
E_g(N\pm \delta N\, ,\, {\cal N}) - E_g(N,\,{\cal N}) =\pm\,
\mu_q^{\pm}\,\delta N\,, \label{endpoints} \\
\delta N/N\rightarrow 0,\quad \delta N/
N^{1/2}\rightarrow\infty\,,
\label{dN}
\end{eqnarray}

\noindent where $E_g$ is the ground-state energy, $\delta N$ is an
arbitrary macroscopically small change in $N$.  The proportionality
coefficients, $\mu_q^{\pm}$, are functions of defects formation energies,
interval $[\mu_q^-,\mu_q^+]$ being the range of the values of the chemical
potential, $\mu$, at which the $q$-crystal exists.

At first glance it would seem that $\delta N$ should be identified with a
number of zero-dimensional defects ("$\pm$defectons"), each of them
arising as one particle is removed from $q$-crystal ($-$defecton) or added
to it ($+$defecton). In such a case $\mu_q^{\pm}$ equals the energy of
$\pm$defecton formation, $\pm\,\epsilon^{\pm}_q$. The crux of the matter
is that this seemingly evident statement is incorrect, as a rule, due to a
{\em coalescence of defectons of the same "sign"}. What this means is
removing from or adding to a $q$-crystal {\em two} particles brings about
a bound state of $-$defectons or $+$defectons ("bidefecton") whose energy
is less than $2\epsilon^{\pm}_q$ (Fig.~\ref{fig1}). We have found out the
coalescence by computer calculations, which were carried out for $v(r)=
\exp(-r/R)/r^\beta$ ($0< R\leq 20, \ 1<\beta<12$) and $\rho \geq 10^{-2}$.
They have shown that the phenomenon occurs at any $q\geq 2$ and for all
$v(r)$ under consideration. The algorithm of the calculations is based
on the dipole description of defectons which is as follows.

Each particle displacement in a $q$-crystal resulting from $\pm$defectons
formation can be considered a "dipole" of a sort. This is a pair
consisting of the particle shifted and the "antiparticle" located at the
host-lattice site left by the particle, the antiparticle "charge" being
equal by magnitude but opposite in sign to that of the particle. A
$\pm$defecton thus comprises a hole ($-$defecton) occupying one of the
host-lattice sites, $\vec r^{\,-}$, or an added particle ($+$defecton)
placed in the $q$-crystal at a free host-lattice site, $\vec r^{\,+}$, and
a system of dipoles, the energy of $\pm$defecton formation having the form

\begin{equation}
\label{epsilon}
\epsilon_{q}^{\pm} =
\min \left( v_{\pm}(\vec r^{\,\scriptscriptstyle\pm}) \pm
\sum_{i=1}^n\widehat\Delta_{\vec \xi_i} v( |\vec r_i - \vec
r^{\,\scriptscriptstyle\pm}|) \, +\, \delta U\,\right)\,,
\end{equation}

\noindent where $\widehat\Delta_{\vec \xi} f(\vec r) \equiv f(\vec r -
\vec \xi) - f(\vec r)$ ($f(r)$ is an arbitrary function); $\vec r_i$ and
$\vec \xi_i$ are the $i$-th dipole antiparticle site vector number and
displacement vector respectively; $\pm\widehat\Delta_{\vec \xi_i} v(|\vec
r_i -\vec r^{\, \scriptscriptstyle\pm}|)$ is the energy of interaction
between the hole /added particle and $i$-th dipole; $n$ is the number of
dipoles; $v_{-}(\vec r^{\,\scriptscriptstyle -})= - 2u_q$ is the energy of
hole formation ($u_q>0$ is the energy of the $q$-crystal per particle),
$v_{+}(\vec r^{\,\scriptscriptstyle +})> 0$ is that of interaction between
the (ideal) $q$-crystal and the added particle; $\delta U>0$ is the
excitation energy of the $q$-crystal produced by the displacements of $n$
particles:

$$\delta U = \sum_{i=1}^n v_q(\vec \xi_i)\,
+\,\frac{1}{2}\sum_{\stackrel{i,k =1}{\scriptscriptstyle i\neq k}}^n
\widehat\Delta_{\vec \xi_i}\widehat\Delta_{\vec \xi_k}
v(|\vec r_i-\vec r_k|)\,.$$

\noindent Here $v_q(\vec \xi) > 0$ is the excitation energy produced by
displacement of a $q$-crystal particle by host-lattice vector $\vec \xi$;
the second term is the energy of the dipole-dipole interaction.
Minimization in (\ref{epsilon}) is performed with respect to variables,
$\vec r_i,\:  \vec \xi_i,\:  n$, and $\vec r^{\,+}$.

To find the energy of formation of a defect with an arbitrary number of
holes/added particles, $m$, (let their host-lattice sites be $\vec
r^{\,\scriptscriptstyle\pm}_{ a}\: (a =1,\ldots, m)$) the first two terms
in Eq.~(\ref{epsilon}) should be replaced by the corresponding sums over
$\vec r^{\,\scriptscriptstyle\pm}_a$, and besides, the additional term,
the energy of mutual holes/added particles repulsion, $V_{\text {rep}} =
\frac{1}{2} \sum_{a\neq b}v(|\vec r^{\,\scriptscriptstyle \pm}_a - \vec
r^{\,\scriptscriptstyle \pm}_b|)$, should be included, among the variables
to be found as a result of the minimization being $\vec
r^{\,\scriptscriptstyle\pm}_a$.

Using the dipole description, the nature of the coalescence can be
clarified. The dipoles of $\pm$defecton are always arranged in such a way
that they are all {\em attracted} to the hole/added particle. The
bidefecton formation causes an {\em extra} attraction between the dipoles
and holes/added particles, as compared with two isolated defectons. The
energy gain exceeds $V_{\text {rep}}$ and thereby results in the
coalescence. This fact is illustrated in Fig.~\ref{fig1}. It should be
noted that for all $q\geq 2$ the structures of the attraction cores of the
bidefectons (the holes and their nearest dipole surroundings) are much the
same. In the case $q=1$ the dipoles do not arise (a $-$defecton is simply
a hole), and hence, the coalescence does not occur.

Formation of a bound state of three $\pm$defectons diminishes the energy
per particle removed/added still further, and so on to the extent of
formation of an infinite {\em strip} of bound $\pm$defectons, which is a
periodic 1D structure with a primitive translation vector $q\vec a$ (see
Fig.~\ref{fig1}). Hence, it is {\em 1D defects}, but not the defectons,
which are bound to arise in a $q$-crystal as a result of a small variation
in $N$.

The simplest 1D defect in a $q$-crystal is a strip of rarefaction or
compression which arises when a $q$-crystal part adjacent to a particle
line with some primitive translation vector, $\vec d$, is shifted as a
whole relative to the other one by a host-lattice translation vector,
$\vec \xi$. Formation of $N_s$ strips results in the change, $\pm\delta
{\cal N}$, in the dimensionless volume of the system: $\pm\delta {\cal
N}=\pm\sigma LN_s$, ($\sigma = 2|\vec \xi\times\vec d|/\sqrt 3,\: L\sim
N^{1/2}$ is the length of a strip). If $N_s /N^{1/2}\rightarrow 0,\;
N_s\rightarrow \infty$, the change in the system energy, $\delta E$, is
proportional to $\delta {\cal N}$:

\begin{equation}
\label{dE}
\begin{array}{c}
\delta E
=\varepsilon(\vec d,\vec\xi)\,\delta {\cal N}\\
\varepsilon(\vec d,\vec \xi) =\sigma^{-1}{\sum_{\vec
r}}^{\prime}\sum_{n=1}^{\infty}\widehat\Delta_{\vec\xi} \,
v(|\vec r - n\vec e|).
\end{array}
\end{equation}

\noindent Here the accent at $\Sigma$ means summation over the half-plane
$\vec r = l\vec d + m\vec e$, ($l =0,\pm 1,\ldots,\: -\infty< m\leq
0 $); \ $\vec e$ is any of the $q$-crystal primitive translation vectors
differing from $\vec d$. Only such strips can be relevant to the
ground-state properties whose vector parameters, $\vec d$ and $\vec \xi$,
realize the minimum of $\delta E$, which equals $E_g(N,{\cal N} - \delta
{\cal N})-E_g(N,{\cal N})$ (rarefaction, $\varepsilon<0$) or $E_g(N,{\cal
N} + \delta {\cal N})-E_g(N,{\cal N})$ (compression, $\varepsilon>0$). We
will call them `$-$strips' or `$+$strips', depending on whether they
result from rarefaction (sign $-$) or compression (sign $+$); their
vectors $\vec d$ ({\em `directors'}) and $\vec \xi$ are denoted here by
$\vec d_q^{\,\pm}$ and $\vec\xi_q^{\pm}$ respectively.  Our computer
calculations based on (\ref{dE}) have shown that the space structure of
$\pm$strips is the same for all $q$ and $v(r)$:

\begin{equation}
\label{univer}
\vec d_q^{\,\pm}= q\vec a_\alpha,\quad \vec\xi_q^{\pm}=\mp\vec
a_\beta,\quad\alpha\neq\beta \,.
\end{equation}

\noindent where vectors $\vec a_{\alpha}$ ($\alpha=1,2,3;\:|\vec
a_{\alpha}| =1$) are the host-lattice primitive translation vectors which
form an angle of $120^{\circ}$ with one another.

As indicated later, $\pm$strips of the same sign {\em repel} each other at
any distances. Hence, the complex strip which result from the
defecton coalescence is bound to dissociate to its constituents,
simple ones.  Noticing also that breaking $\pm$strips into fragments
shifted relative to one another increases inevitably the energy of the
system (this physically evident statement is completely confirmed by our
computer calculations), we can conclude that it is just the $\pm$strips
which are the above-mentioned 1D defects.

As follows from simple thermodynamic considerations, the coefficients
$\mu_q^{\pm}$ in Eq.~(\ref{endpoints}) are related to the energy of
$\pm$strip formation, $\varepsilon_q^{\pm} = |\varepsilon(\vec d_q^{\pm},
\vec\xi_q^{\pm})|$, by the expression

\begin{equation}
\label{mu_strips}
\mu_q^{\,\pm} = q^2\varepsilon_q^{\pm} + u_q\,.
\end{equation}

\noindent Lest there be a contradiction to the fact of the defecton
coalescence, the energies of $\pm$defectons formation are bound to be
external to the interval $[\mu_q^{\,-}, \mu_q^{\,+}]$. This can be
established by a rather simple reasoning for $v(r)$ which goes to zero
over distances $R\ll\overline r\sim q$. As $|\vec\xi_q^{\pm}|=1$, from
Eq.~(\ref{dE}) we have $\varepsilon_q^{\pm}\sim (qR)^{-1}u_q$. On the
other hand, $|\epsilon_-|$ is comparable with $u_q$, and hence,
$\mu_q^{\,-}\sim (q/R)u_q \gg |\epsilon_-|$. To evaluate $\epsilon_q^{+}$
one should take into account that $+$defecton formation significantly
decreases the least of the interparticle distances, $r_{min}$, as compared
with $\overline r$.  Therefore, $\epsilon_q^+ \sim v(r_{min})\gg \mu
_q^{\,+}\sim (q/R) v(\overline r)$. To make sure that the inequalities,
$|\epsilon_q^-| < \mu_q^{\,-}< \mu_q^{\,+}<\epsilon_q^+$, hold
irrespective of a form of $v(r)$ the computer calculations of
$\mu_q^{\pm}$ based on Eq.~(\ref{epsilon}), (\ref{dE}), and
(\ref{mu_strips}) have been carried out in parallel to the computer
studies of the coalescence. They have confirmed that the above
inequalities are really the case for all $v(r)$ under consideration and
for all $q\geq 2$.

The above reasoning shows that at $|\rho - \rho_q|\ll 1$ the 2D LSLRIR
{\em ground-state} space structure is a kind of a dilute solid solution of
$-$strips or $+$strips (depending on the sign of $\rho - \rho_q$) in the
$q$-crystal, $\pm$strips sharing their primitive translation vector,
$q\vec a_\alpha$, with `unperturbed' $q$-crystal strips. It should be
noted that due to the degeneration caused by the hexagonal symmetry of the
$q$-crystal (see Eq.~(\ref{univer})) at a given $\rho$ superstructures can
exist which comprise strips with different $\vec d_q^{\pm}$. However,
these cannot be the ground state, as intersections of the strips
inevitably result in an extra increase in the system energy. It can also
be shown that the same holds for the mixtures of $\pm$strips with the same
director but different displacement vectors, $\vec \xi_q^{\pm}= \mp \vec
a_\beta$.  To find the arrangement of the $\pm$strips with a given
director and a given displacement vector one can consider the 2D LSLRIR as
an 1D LSLRIR whose `particles' are {\em particle lines} of the 2D system,

\begin{equation}
\label{lines}
\vec r = kq\vec a_\alpha + l\,\vec a_\beta\quad (k=0,\pm 1,\pm 2,\ldots;
\:\:\alpha\neq\beta),
\end{equation}

\noindent with integers $l$ as `coordinates'. From this point of view a
$\pm$strip is a pair of lines (\ref{lines}) which are shifted relative to
one another by vector $(q\mp 1)\vec a_\beta$, while the relative
displacement vector of the lines which form an unperturbed strip is $q\vec
a_\beta$.

The line arrangement is governed by the pair potential of line-line
interaction,

\begin{equation}
\label{vll}
V_q(l_2-l_1)= \sum_{k=-\infty}^\infty
v\,(|kq\vec a_\alpha + (l_2-l_1)\,\vec a_\beta|)\,,\quad (\alpha\neq\beta),
\end{equation}

\noindent where $l_1, l_2$ are `coordinates' of interacting lines. Simple
computer calculations show that due to convexity of $v(r)$ the pair
potential (\ref{vll}) is positive and monotonically decreasing function of
$l$ which also meets the convexity condition,
$F_q(l)\equiv V_q(l)-2V_q(l+1)+V_q(l+2)> 0$, over the whole range $l>1$
and for all $q\geq 2$. It follows herefrom that the arrangement of the
lines (\ref{lines}), at least in some vicinity of $\rho_q$, is the same as
that of the 1D LSLRIR particles, and hence, it obeys the universal Hubbard
algorithm \cite{Hubbard,Bak,Synay}

\begin{mathletters}
\label{arr}
\begin{equation}
\label{arr_lines}
l_m =[\overline l\, m],\quad  \overline l =\vartheta/ q
\end{equation}

\noindent where $\overline l$ is the mean line separation, $\vartheta=
1/\rho$, $l_m$ is the `coordinate' of $m$-th line.  The algorithm
determining the arrangement of $\pm$strips can be found directly from
Eq.~(\ref{arr_lines}):

\begin{equation}
\label{arr_strips}
n_m = [m/c_s^{\scriptscriptstyle\pm}]; \quad
q(q \mp c_s^{\scriptscriptstyle\pm}) = \vartheta \quad
0\leq c_s^{\scriptscriptstyle\pm}\leq 1.
\end{equation}
\end{mathletters}

\noindent Here $m$ enumerates the $\pm$strips, $n_m$ is the number of the
particle lines between $m$-th $\pm$strip and $\pm$strip with $m=0$;
$c_s^{\scriptscriptstyle\pm} =N_s^{\scriptscriptstyle\pm}/N_l$ is the
concentration of the $\pm$strips ($N_s^{\scriptscriptstyle\pm}$ and $N_l$
are total numbers of the $\pm$strips and the particle lines
respectively).

The 2D LSLRIR ground-state space structures (\ref{arr_strips}) are stable
not only in some vicinity of $\vartheta = q^2$, but also over the whole
range of $\pm$strip concentrations, $0\leq c_s^{\scriptscriptstyle\pm}\leq
1$. This is due to the following closely correlated facts, which we (using
computer calculations) have found to take place for any $v(r)$ under
consideration:  a)~zero-dimensional defects (defectons) of a structure
described by Eq.~(\ref{arr}) coalesce irrespective of a
$c_s^{\scriptscriptstyle\pm}$ value; b)~therefore, an infinitesimally small
variation in $c_s^{\scriptscriptstyle\pm}$ produces only 1D defects in the
structure; c)~their primitive translation vectors turn out to be equal
to $q\vec a_\alpha$ for any $c_s^{\scriptscriptstyle\pm}$.

At $c_s^{\scriptscriptstyle -}=1$ ($\vartheta = q(q + 1)$) algorithm
(\ref{arr}) determines the stable simple crystal with primitive
translation vectors $q\vec a_\alpha,\:  (q + 1)\vec a_\beta \:
(\alpha\neq\beta)$. This can also be considered to be the structure
arising at $c_s^{\scriptscriptstyle +}=1$ as a result of compression of
the $q+1$-crystal. Hence, universal algorithm (\ref{arr}) $(q=
2,3,\ldots)$ covers the whole range $2\leq\vartheta<\infty$, expressions
(\ref{arr}) holding over the intervals $q(q-1)\leq\vartheta\leq q(q+1)$.
For each interval there is its own set of directors, $q\vec a_\alpha$.

The only exception, as our computer calculations have shown, is a vicinity
of $\vartheta= 2$ where the above algorithm fails in the case of
slow-diminishing pair potentials $v(r)$. Particularly, if $v(r) =
\exp(r/R)/r$ or $v(r) = 1/r^\gamma$, this occurs for $R\geq R_{0}\approx
0.38$ or $\gamma\leq \gamma_{0}\approx 4.3$ respectively. The matter is
that for $v(r)$ of such a type the coalescence of $\pm$defectons of the
simple crystal with $\vartheta = 2$ (its primitive translation vectors are
$\vec a_\alpha, 2\vec a_\beta,\: (\alpha \neq \beta)$) leads to formation
of 1D defects the primitive translation vector of which, $4\vec a_\alpha$,
is {\em doubled} in length as compared with director corresponding to
$q=2$.  These defects cannot disassociate into simple strips of the above
type. Our preliminary results suggest that on interval the
$2\leq\vartheta\leq 4$ there exists a critical point (its position depends
on a form of $v(r)$), at which the 2D LSLRIR ground state undergoes a
first-order structure transition with doubling of a spacing, $2\vec
a_\alpha \rightarrow 4\vec a_\alpha $. We are going to carry out a
detailed study of the situation in a separate paper.

The 2D LSLRIR ground-state dependence of the chemical potential, $\mu$, on
$\rho$, just as it takes place for the 1D LSLRIR \cite{Hubbard,Bak,Synay},
is a devil staircase whose steps occur at all rational $\vartheta$. Taking
into account Eq.~(\ref{arr_lines}), one can represent $\vartheta$ in a
form $\vartheta =q\overline l$ ($q=2,3,\ldots$), where $\overline l= Q/P$,
and $P,Q$ are any coprime integers meeting conditions $q-1\leq Q/P\leq
q+1$. Thus, the 2D LSLRIR ground state is a set of crystals with $P$
particles per cell and primitive translation vectors $q\vec a_\alpha,
Q\vec a_\beta, \: \alpha\neq\beta$, the crystals with one particle per
cell having $\vartheta = q^2$ or $q(q+1)$.  Using results of the 1D LSLRIR
theory \cite{Bak,Synay}, we have found the widths of the devil-staircase
intervals, $\Delta\mu$, to be related to the line pair potential
(\ref{vll}) by an expression

\begin{equation}
\label{width}
\Delta\mu  = Q\sum_{m=1}^{\infty} mF_q(Qm -1) \, >0
\end{equation}

\noindent At a given $q\geq 2$ the expression holds over the interval
$q(q-1)\leq\vartheta\leq q(q+1)$. It should be noted that $\Delta\mu$,
similarly to the 1D case, does not depend on the number of particles per
cell, $P$. In the case of $q$-crystal ($Q=q,\:P=1$) expression
(\ref{width}) coincides with $\mu_q^+-\mu_q^- $, (see
Eq.~(\ref{mu_strips}).

The above theory holds over the interval $1 \geq \rho\geq 1/2$ if one
applies it to the holes (empty host-lattice sites) as the particles, the
filling factor of holes, $\rho_h$, being equal to $1-\rho$. It is evident
that the hole space structure corresponding to a given $\rho_h$ is
equivalent to the particle one with $\rho = 1 - \rho_h$.

In essence, the effective lowering of dimension and the coalescence of
defectons underlying it are caused by discreteness of the system under
consideration. Therefore, the phenomena should be expected to exist
irrespective of the geometry of a host lattice. Our preliminary studies
have confirmed this suggestion, yet they have shown that for an arbitrary
host lattice the structure of the 2D LSLRIR ground state turns out to be
more sophisticated than in the case of a triangular host lattice. We also
have revealed that at non-zero temperatures the lowering of dimension
leads to an interplay of a strongly correlated liquid of the thermally
fractured 1D defects and an ideal gas of defectons mentioned. This results
in a very interesting low-temperature thermodynamics of the 2D LSLRIR. We
are going to publish the results concerning these issues in the near
future.

We would like to thank M.Pepper for bringing his experimental results
to our attention.

\begin{figure}
\caption{The coalescence of $\quad -$defectons in $q$-crystal with $q=4$.
Here $\circ$ denotes host-lattice sites,
$\protect\large\bullet$ - particles; $\odot$ - antiparticles; $\otimes$ -
holes; $\rightarrow$ - dipoles; the dotted lines mark off a single
defecton, a bidefecton. The attraction core of the bidefecton consists of
dipoles $D_1$, $D_2$, and holes $H_1$, $H_2$. The total energy of
attraction $H_1 \leftrightarrow D_2$ and $H_2\leftrightarrow D_1$ exceeds
the energy of repulsion $H_1\leftrightarrow H_2$. With an increase of
number of removed particles the strip of bound defectons arises, the
holes of the strip stringing out along a particle line passing
through the sites $H_1$, $H_2$.}
\label{fig1}
\end{figure}


\begin{references}
\bibitem{Pokrovsky} V.~L.~Pokrovsky, A.~L.~Talapov,
Theory of incommensurate crystals, N.Y., Harward Acad., 1984;
I.~F.~Ljuksutov,  A.~G.~Naumovets, V.~L.~Pokrovsky
Two-Dimensional Crystals, Naukova Dumka, Kiev, 1988.
\bibitem{FNT93_UFN95} A.A.Slutskin and L.Yu.Gorelik, Low Temp.\ Phys.\
{\bf 19}(11), 852-864 (1993). A.A.Slutskin,
Physics, uspekhi, June 01 1995, v.38, No 6, p.669.
\bibitem{Heywang} W. Heywang, Amorfe und Policrystalline Halbleiter,
Springer Verlag, 1984.
A.A. Slutskin, V.I. Makarov.
in: Proc. of Electroceramics 5, September 1996, Aveiro (Portugal),
p.267.
\bibitem{Pepper} M Pepper. J. Phys. C: Solid State Phys., {\bf 12}, L617
(1979)
\bibitem{Hubbard} Hubbard J., Phys. Rev. B {\bf 17}, 494 (1978).
\bibitem{Bak} P.~Bak, R.~Bruinsma. Phys. Rev. Lett. {\bf 49}(4), 249
(1982)
\bibitem{Synay}Ya.~G.~Sinai, S.~Ye.~Burkov. Uspekhy Matematicheskikh Nauk,
{\bf38}(4), 205 (1983)
\bibitem{SS} V.V.Slavin, A.A.Slutskin, Phys. Rev.B v.54(11),
p.8095-8100 (1996).
\end{references}
\end{document}